\documentclass[aps,pre,twocolumn]{revtex4-2}
\usepackage{amssymb}
\usepackage{amsmath,bm}
\usepackage{graphicx,color}
\usepackage{bbm}
\usepackage{appendix}
\newcommand{\B}[1]{{\bm{#1}}}

\usepackage{physics}
\usepackage{enumitem}

\begin{document}
\title{Generic Mechanism for Remote Triggering of Earthquakes}
\author{Harish Charan$^1$, Anna Pomyalov$^1$ and Itamar Procaccia$^{1,2}$}
\affiliation{$^1$Dept. of Chemical Physics, The Weizmann Institute of
	Science, Rehovot 76100, Israel.\\  $^2$Center for OPTical IMagery Analysis and Learning, Northwestern Polytechnical University, Xi'an, 710072 China. }

\begin{abstract}
``Remote triggering" refers to the inducement of earthquakes by weak perturbations that emanate from 
far away sources, typically intense earthquakes that happen at much larger distances than their nearby
aftershocks, sometimes even around the globe. Here, we propose a mechanism for this phenomenon; the proposed mechanism
is generic, resulting from the breaking of Hamiltonian symmetry due to the existence of
friction. We allow a transition from static to dynamic friction. {\em Linearly stable} stressed systems display giant sensitivity to small perturbations
of arbitrary frequency (without a need for resonance), which trigger an instability with exponential oscillatory growth. Once nonlinear effects kick in, the blow up in mean-square displacements can reach 15-20 orders of magnitude. 
Analytic and numerical results of the proposed model are presented and discussed.   

\end{abstract}
\maketitle

\section{Introduction}

It was said that when the magnitude 7.3 Landers earthquake struck in 1992 in the desert north of Palm
Springs, the earthquake map of the state of California lit up like a Christmas tree \cite{05Hou}.
Since the early nineties of the last century there accumulated growing evidence that remote earthquakes can trigger subsequent large earthquakes with epicenters far
away from the original one, occasionally even far around the world \cite{18OMGB}. While it is obvious that seismic waves propagate in the earth crust, intense ones are typically highly damped, and only long wavelength perturbations, which
are relatively weak, can reach long distances. It is then natural to ask, what might be the mechanism for
the amplification of weak perturbations that could be behind this so called ``remote triggering" \cite{13ESKA}. In seeking such a mechanism, one must bear in mind a few basic requirements. Firstly, one needs to think about a system that is already under the action of strain and stress, but which is linearly stable, i.e. it does not erupt in the absence of an external perturbation. Secondly, the system has to be sensitive to weak perturbations, since, as said, only weak, long wavelength perturbations are playing a role in remote triggering. Thirdly, the frequency of the weak perturbation should not have any particular relation to the natural frequencies of the perturbed system - a generic mechanism should allow a giant response to arbitrary frequencies of the incoming perturbing wave. In other words, we are not looking for a resonance effect. Finally, the mechanism should be reasonably generic in the sense that it would not depend on very special characteristics of the perturbed system. 

In this paper we propose a model that satisfies all these requirements. To be precise, we do not
claim that it is the only possible model for remote triggering, but we would like to offer it as a reasonable candidate for further examination and testing. We argue that the wanted properties
result from the breaking of Hamiltonian symmetry, bringing about instabilities that are not present in systems whose dynamics is derivable from a Hamiltonian. Of course, one can break Hamiltonian symmetry in many ways, and here we stress the role of friction. 

The structure of this paper is as follows: the model is described in detail in Sect.~\ref{model}. The instabilities of this model and the giant sensitivity to weak perturbations
are discussed in details in Sect.~\ref{properties}. In Sect.~\ref{damping}, we offer a critique of the model and what might destroy its relevance to seismic applications. The final section \ref{summary} offers a summary of the paper and some conclusions.

\section{The model}
\label{model}

\subsection{The setup}

\begin{figure}
	\hskip -0.5 cm
	\includegraphics[scale=0.28]{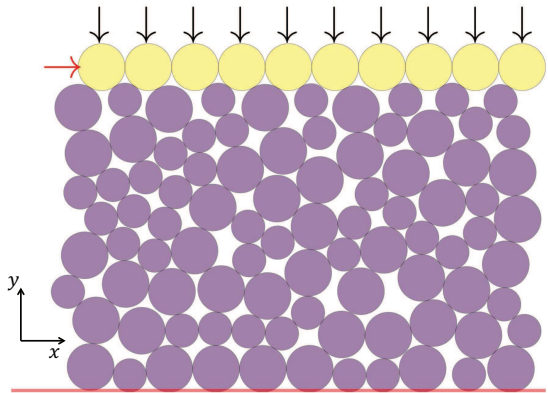}
	\caption{The model: an amorphous array of $N$ disks of two sizes [here, $N$ = 100], compressed between a flat horizontal wall, here shown in red color line, and an upper boundary made of disks shown in yellow. The lower wall is fixed and the upper
		wall is movable. The entire system is under the action of constant external force in y-direction $F_{y}$, 
		while the disks forming the upper wall are acted upon by an additional force in x-direction $F_{x}$.}
	\label{modelfig}
\end{figure}

A very popular model for the eruption of earthquakes is the Burridge–Knopoff spring-block model \cite{67BK}. This model employs many blocks interconnected by elastic springs with a spring stiffness coefficient. The blocks are also elastically coupled with a spring stiffness coefficient to a rigid plate moving at a constant velocity, and pulled over a rough surface described by some friction law. The interface between the blocks and the rough surface can be considered an analogue for a 1-D earthquake fault \cite{91CLST}. For our purposes this model is not ideal, since we want to understand {\em what initiates} the fast relative movements of the fault plates. We are looking for a mechanism that induces the {\em beginning of motion} in systems that are linearly stable while being under strain.
In other words, we are interested in a mechanically stable situation that gets destabilized by weak perturbations, leading to a major event. Moreover, we respect the fact that faults are usually filled up with a gouge which is not faithfully represented by blocks and springs. We therefore propose the model shown in Fig.~\ref{modelfig}.
The model consists of $N$ disks of mass $m$ in a box of size $L^2$, having two different 
radii $R_1=0.5$ or $R_2= 0.7$ respectively. This is done to avoid forming a periodic array, keeping an amorphous structure at all times. Two disks exert forces on each other only when
they overlap. There are two kinds of interaction forces between disks. The first is a normal Hertzian force which depends upon the overlap between two disks when they come into
contact. This always pushes disks apart. The Hertzian force exerted by disk $j$ at {\bf r}$_{j}$ on a disk 
$i$ at {\bf r}$_{i}$ is given as:
\begin{equation}
	{\bf F}_{ij}^{(n)} = k_n \delta_{ij}^{3/2}\hat r_{ij} \,
\end{equation}
where $k_n$ is an elastic constant that depends upon the material properties, and 
\begin{eqnarray}
	\delta_{ij} &=& (R_{i} + R_{j}) -r_{ij}\ , \nonumber\\
	 {\bf r}_{ij}&=& {\bf r}_i-{\bf r}_j \ , \quad
\hat r_{ij} \equiv {\bf r}_{ij}/r_{ij} \ . 
\label{defs}
\end{eqnarray}

The second force is a tangential force $\B F_{ij}^{(t)}$ which depends on the tangential displacement $t_{ij}$ of two disks
when they overlap. This is due to friction, always opposing the motion in tangential directions. The determination of the $t_{ij}$ is dynamical, and cannot be read from a given snapshot, although the tangential forces {\em can} be determined in principle, cf. Ref.~\cite{16GPP}. The tangential forces 
depend on the angular dynamics of the disks, adding a degree of freedom $\theta_i$ for each
disk, where by convention $\theta_i=\theta_j=0$ when the disks $i$ and $j$ form contact for the first time. The calculation of the tangential displacements is detailed in Appendix~\ref{tang}.

The tangential force is assumed to be bounded by the Coulomb law,
\begin{equation}
	\B F_{ij}^{(t)}	\le \mu(v)\B F_{ij}^{(n)}
	\label{coulomb}
	\end{equation}
where $\mu(v)$ is a friction coefficient. As long as the relative velocities between grains do not exceed a given threshold, $\mu(v)$ is a static coefficient $\mu_0$. Once such velocities appear, a dynamic friction coefficient replaces the static one in Eq.~\ref{coulomb}. The velocity dependence of the dynamic friction coefficient is discussed in the next subsection.  In our model, the forces are smooth functions of the displacements, with a smooth first derivative; in light
of the bound Eq.~(\ref{coulomb}) we represent the tangential interaction as a 
Mindlin force exerted by disk $j$ at {\bf r}$_{j}$ on a disk $i$ at {\bf r}$_{i}$:
\begin{equation}
	{\bf F}_{ij}^{(t)}\! =\! -k_t\delta_{ij}^{1/2}\!\left[1\!+\!\frac{t_{ij}}{t^*_{ij}} \!-\!\left(\frac{t_{ij}}{t^*_{ij}}\right)^2\right]\!t_{ij} \!\hat t_{ij} \ , \quad t^*_{ij}\! \equiv\! \mu \frac{k_n}{k_t} \delta_{ij} \ ,
	\label{Ft}
\end{equation}
with $k_t$ being a material parameter. We note that the tangential force depends on the square-root of the normal overlap
$\delta_{ij}$. This is physical, since the frictional force should increase with 
the increased contact between the disk. The chosen form of the tangential force Eq.~(\ref{Ft}) guarantees
that it reaches the Coulomb limit with smooth first and second derivatives. In this work we use units of mass, length and time $m$, $2R_1$ and $\sqrt{m(2R_1)^{1/2} k_n^{-1}}$ respectively.

Besides the fact that we smooth out the derivatives of the tangential forces at the Coulomb limit, the model forces are quite traditional, following the frequently used Cundall-Strack model of 
frictional disks \cite{79CS}. For our purposes, the important property of this model is that the forces are not derivable from a Hamiltonian. The fact that the normal force does not depend on the
tangential one, but the latter does depend on the former, precludes any model Hamiltonian from
which these can be derived. This broken Hamiltonian symmetry is all-important for our discussion.
Otherwise the precise form of the model forces is less important. 

The assembly of disks is compressed between two boundaries, the lower one being
a flat wall and the upper one being a line of disks. A uniform (and constant) force $- F_y \hat y$ is applied on all the upper wall disks in the downward direction. In addition, we apply a force $F_x \hat x$
in the positive $x$ direction on the disks of the upper layer. This force will be used
as a control parameter. The forces between the boundaries and the disks in contact with them
are the same normal and tangential forces as between the disks themselves. 

\subsection{Dynamic Friction Coefficient}

\begin{figure}
	\includegraphics[scale=0.5]{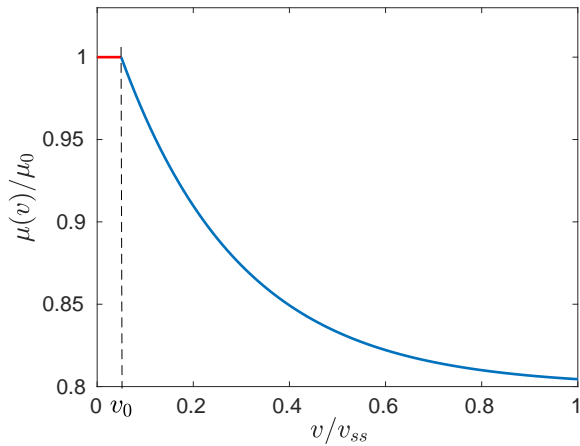}
	\caption{The velocity dependence of the friction coefficient appearing in Eq.~\ref{coulomb}. When the velocity of the upper layer is  smaller than a predefined threshold value $v_0$, the friction coefficient is constant. For $v>v_0$, $\mu(v)$ gradually decreases toward its steady-slide value.}
	\label{vdep}
\end{figure}

The dependence of the dynamic friction coefficient on the relative velocity of bodies in contact is a complex issue that is not fully understood \cite{97NKG,98NKBG,04DiToro,Dieterich78,AharonovScholz18}. A short review of some of the experimental knowledge about this issue is presented in Appendix ~\ref{dynfric}. In general, it is stated that once the relative velocity exceeds some threshold value $v_0$, the dynamical friction coefficient is expected to become smaller than the static one, reflecting the fact that the creation of contacts is a plastic process that takes time, time that is not available when the bodies are in motion \cite{98NKBG}. At some larger velocity $v_{\rm ss}$, a steady-slide motion is reached with $\mu_{\rm ss}<\mu_0$, and see Appendix \ref{dynfric} for details. In our simulations, we  use a simple model-law for the dependence of the friction coefficient on the velocity of the upper layer. The justification of this simple form is provided in the said Appendix \ref{dynfric}. Here we provide the velocity dependence of the dynamic friction coefficient as used in the simulations
\begin{equation}\label{muv}
\mu(\tilde v) =\mu_0\Big(B e^{-w(\tilde v - \tilde v_0)}+ \mu\sb{ss}\Big)\, ,
\end{equation}
where $\tilde v=v/v_{\rm ss}$ and $B$ and $w$ are  fitting parameters, discussed in  Appendix~\ref{dynfric}.
The form of this law is shown in Fig.~\ref{vdep}. 

In our simulations we use the parameters $B=0.2$, $w=0.4$, $\mu_0=1$, $\mu_{\rm ss}=0.8$ and the threshold values $\tilde v_0$ are defined by the undamped Newtonian dynamics of the stable configuration.
\section{Instabilities and giant sensitivity to weak perturbations}
\label{properties}

\subsection{Dynamics}

To take into account the angular and positional dynamics we employ an
extended set of coordinates $\B q_i=\{\B r_i, \B \theta_i\}$. Their equations
of motion are Newtonian, i.e.
\begin{eqnarray}
	\label{Newton2a}
	m_{i}\frac{d^2 \B r_{i}}{dt^2}&=&{\B F}_i(\B q_1,\B q_2,\cdots,\B q_N)\ , \\
	\label{Newton2b}
	I_{i}\frac{d^2 \B \theta_{i}}{dt^2}&=&{\B T}_i(\B q_1,\B q_2,\cdots,\B q_N)\ ,
\end{eqnarray}
where $m_i$ are masses for the coordinates, $I_i$'s are moments of inertia for the angles. ${\B F}_i$'s are resultant forces (sum of all normal and tangential binary forces on each disk), and ${\B T}_i$'s are torque. Using the smoothed out forces Eq.~\eqref{Ft}, this allows to define the stability matrix $\B J$ which is an operator obtained from the derivatives of the force $\B F_i$ and the torque $\B T_i$ on each particle with respect to the coordinates. In other words
\begin{equation}
	J_{ij}^{\alpha\xi} \equiv \frac{\partial \tilde F^\alpha_i}{\partial q_j^\xi}\ , \quad \tilde{ \B F}_i \equiv \sum_j \tilde{\B F}_{ij} \ ,
\end{equation}
where $\B q_j$ stands for either a spatial position or a tangential coordinate, and $\tilde {\B F}_i$ stands for either a force or a torque. Since in the
usual case $\B F_i =-\partial U/\partial \B r_i$ we see that the operator $\B J$ is an analog
of the Hessian even when a Hamiltonian description is lacking.
But with a huge difference: $\B J$ is not a symmetric operator.
Being real, it can possess pairs of complex eigenvalues. The solution of the linearized stability problem is $\exp{i\omega t}$ where the frequency $\omega$
is related to the eigenvalue $\lambda$ of $\B J$ according to
\begin{equation}
	\omega=\pm\sqrt{\lambda} \ .
\end{equation}
Thus when the eigenvalue becomes complex, with real and imaginary parts, $\lambda=\lambda_{\rm R} \pm i\lambda_{\rm Im}$, there are four frequencies that are complex as well.
When these appear, the system will exhibit
oscillatory instabilities, since one of each complex pair will cause an oscillatory exponential divergence of any perturbation, and the other an oscillatory exponential decay. For the parameters of the simulations discussed below the typical frequency associated with the oscillatory instability near onset is of the order of $\omega\approx \times 10^{-5}$.

The actual calculation of the operator $\B J$ is somewhat cumbersome but
 conceptually straightforward. A detailed calculation
for the present model is presented in Appendix ~\ref{jacobian}.

\subsection{Protocol of simulations}
To initiate the simulations, we start with loosely packed disks in a square box and let them settle under the influence of the contact forces and the constant force $-F_y \hat y$,
until the maximum force between the disks drops below a small predefined value of $f_{\rm min}=10^{-15}$. The volume fraction of this stable system is about $\phi=0.8171$, see Fig\ref{modelfig}.

We then apply to all disks of the upper layer a constant horizontal force $F_x$ pointing toward the positive $x$-direction. This force is increased  by small increments $\delta F_x=0.0001$ and the system is allowed to reach a new equilibrium state with all contact forces smaller than $f_{\rm min}$. For each such equilibrium state,  we  calculate the  operator $\B J$ and its eigenvalues.
At a certain value $F_x$, a pair of complex eigenvalues bifurcates, indicating that the system becomes unstable to external perturbations.

The simulations are carried out using the discrete element method developed by Cundall and Strack~\cite{79CS}, as
implemented in LAMMPS with modifications to include Eq.\eqref{Ft}. The equilibrium state is obtained using over-damped dynamics, while in the analysis of the instabilities the viscous damping is turned off.

\begin{figure}
	\includegraphics[scale=0.45]{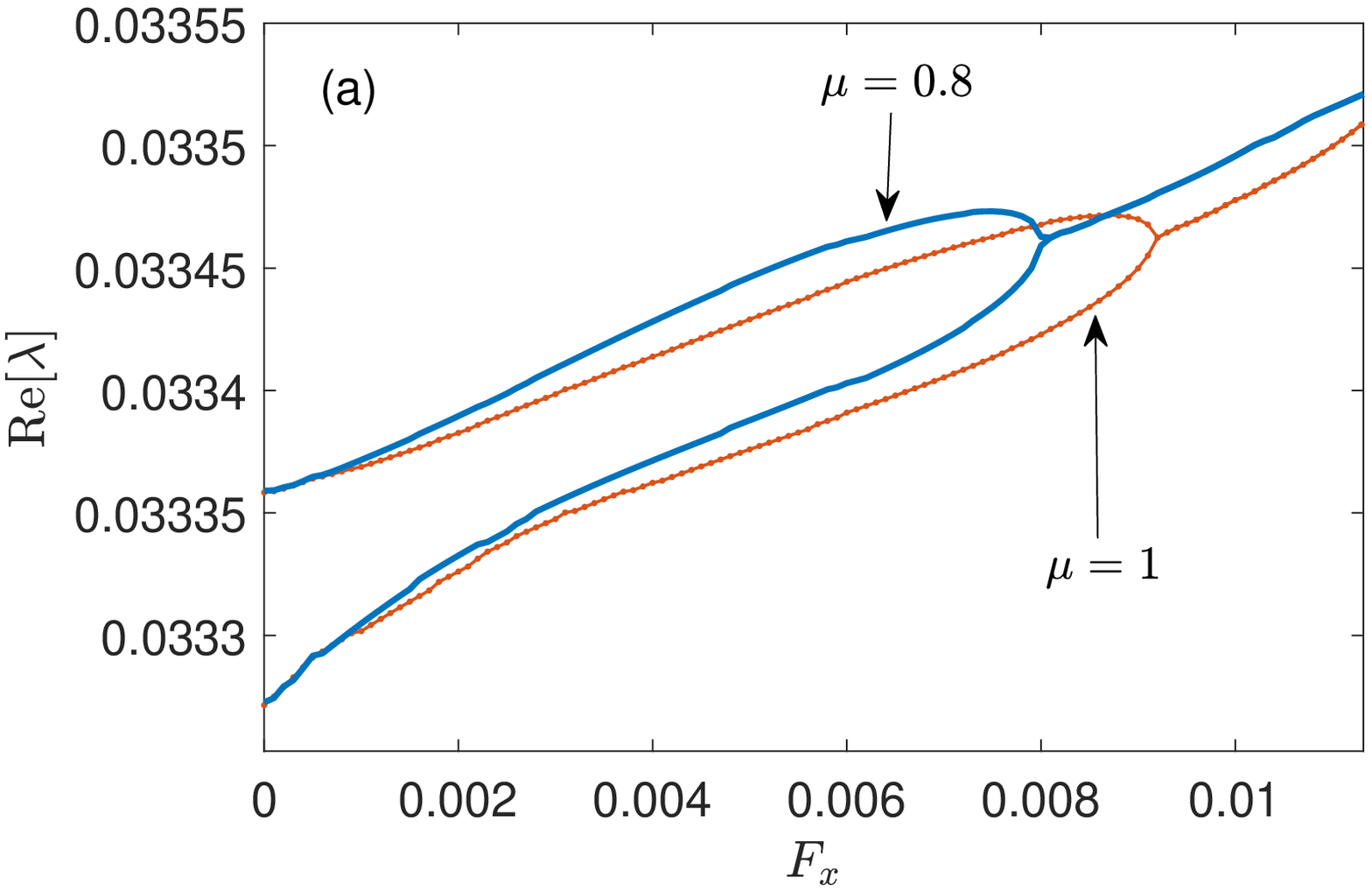}
	\includegraphics[scale=0.44]{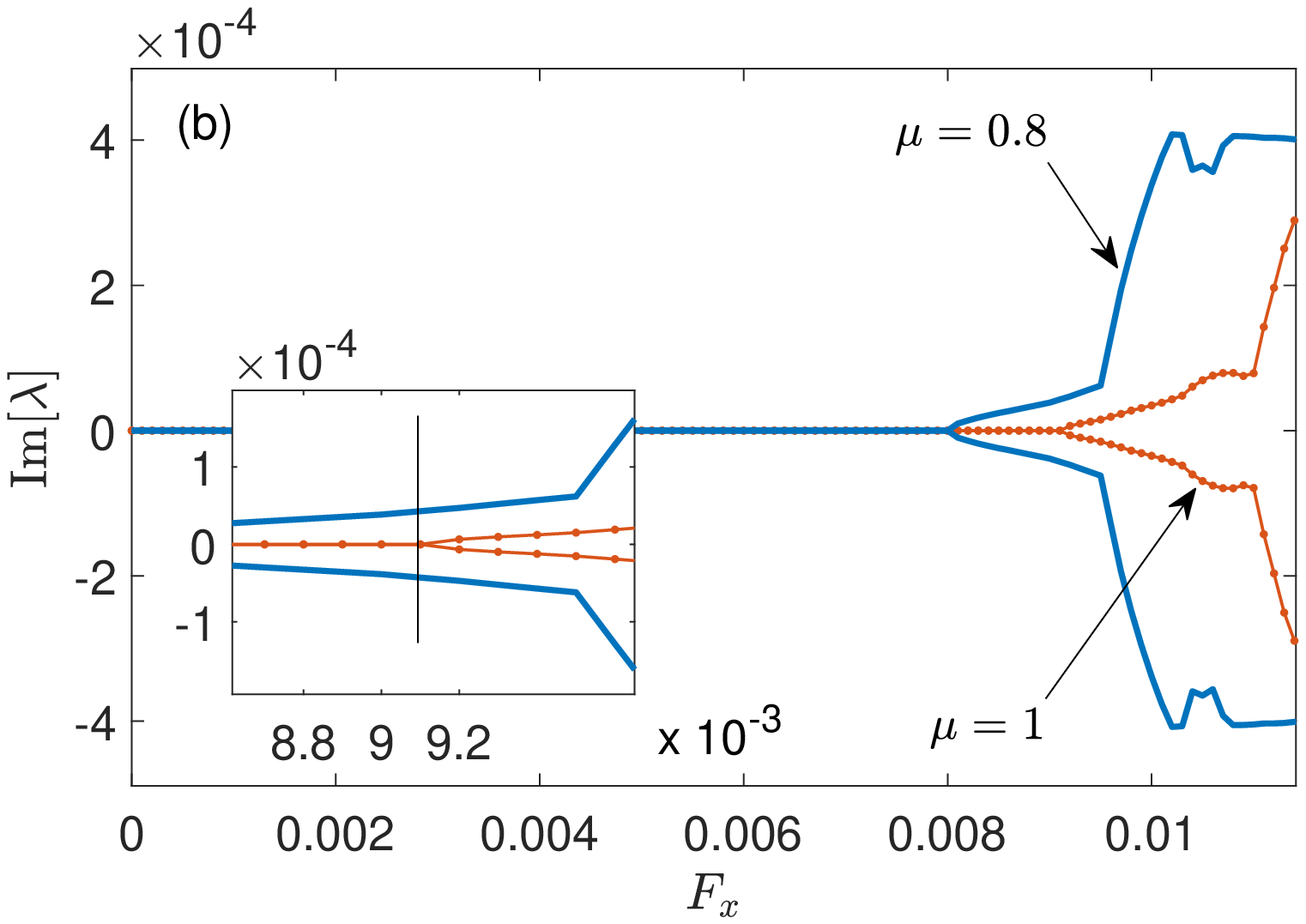}
	\caption{{(a)} A pair of real eigenvalues coalesces upon increasing the horizontal force $F_x$ to the value $F_x^c$. (b) At the critical value $F_x^c$, a pair of imaginary parts bifurcates. These bifurcations occur at smaller critical values for smaller friction coefficient  $\mu$. The real  and imaginary components of the eigenvalues are shown for $\mu=0.8$ ( blue lines) and $\mu=1$ (red lines with dots). The inset in (b) magnifies the region near the onset of bifurcation for $\mu=1$. The vertical black line corresponds to $F_x=0.0091$.  The parameters of this simulation are $k_n=2\times 10^5, k_t=2k_n/7 , F_y=3$.}.
	\label{fdep}
\end{figure}

To study the influence of the friction coefficient $\mu$ on the system's stability, the above protocol was repeated for different values of $\mu$.  As illustrated in Fig~\ref{fdep}, for smaller $\mu$, the birth of the complex eigenvalues occurs at smaller values of $F_x$.

\subsection{Instabilities} 

\subsubsection{Oscillatory Instability}

\begin{figure}
	\hskip -0.5 cm
	\includegraphics[scale=0.60]{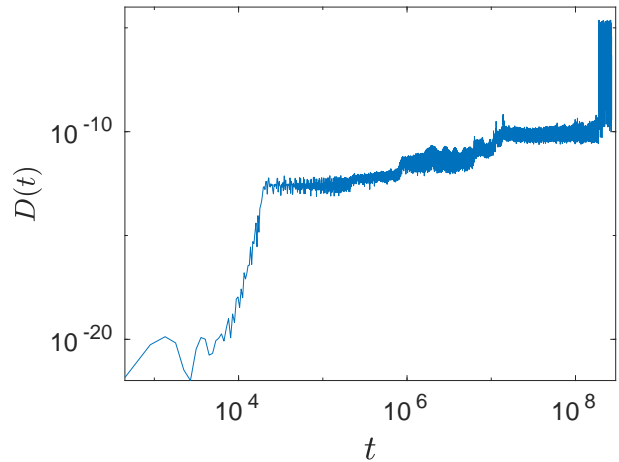}
	\caption{The time dependence of the MSD in the linearly unstable regime. The unstable dynamics increase the numerical noise by some 15 orders of magnitude, where nonlinear effects dominate.}
	\label{MSD}
	\end{figure}

The appearance of the oscillatory instability of interest can be observed by following the eigenvalues of the Jacobian matrix as a function of $F_x$. At small
values of $F_x$ all the eigenvalues are real and positive. At a critical value of $F_x=F_x^c$ two real eigenvalues coalesce, cf. Fig.~\ref{fdep} panel (a). The critical value depends of course on the friction coefficient $\mu$, decreasing as $\mu$ decreases. Simultaneous with the coalescence of the real parts of the eigenvalues, a pair of imaginary parts bifurcates, cf. Fig,~\ref{fdep} panel (b). As explained above, the appearance of imaginary eigenvalues inevitably leads to an oscillatory instability. Once we are in the unstable domain $F_x>F_x^c$, the system
will spontaneously magnify any infinitesimal perturbation, in the present case
even numerical noise. The development of the instability can be monitored by
following the mean-square-displacement (MSD) from the state of mechanical equilibrium. Denoting the MSD as $D(t)$ we compute
\begin{equation}
	D(t)\equiv \frac{1}{N} \sum_i^N [\Delta x_i^2(t) + \Delta y_i^2(t)] \ .
	\label{defMSD}
\end{equation}
The time dependence of this quantity for $\mu=1$ and $F_x=0.012$  is shown in Fig.~\ref{MSD}. 
We see that $D(t)$ increases (starting from just numerical noise) by some 15 orders of magnitude, signaling 
a very violent response of the system to a very small increase in $F_x$. 

\subsubsection{Sensitivity to small perturbations} 

Having demonstrated the instability for $F_x>F_x^c$, we are interested now
in a linearly stable system at $F_x<F_x^c$ which is subjected to a small perturbation. It is important to realize that the type of instability discussed
here renders the linearly stable system highly sensitive to minute perturbations. The theory behind this important feature was discussed in detail in Ref.~\cite{20CGPS}.
Here we summarize the essential observations.  Choosing appropriate units
of time, the normal form for our instability can be written as
\begin{equation}
	\partial_{t}^{2}\begin{pmatrix}x\\
		y
	\end{pmatrix}=- \B J\begin{pmatrix}x\\
		y
	\end{pmatrix} = -
	\begin{pmatrix}1-\delta & \eta\\
		-\eta & 1+\delta
	\end{pmatrix}\begin{pmatrix}x\\
		y
	\end{pmatrix}\label{eq:hom}
\end{equation}
with $1>\delta$.

Substituting $\left(x,y\right)=\left(X,Y\right)e^{i\omega t}$ with constant $X,Y$
we find that the eigenfrequencies are obtained as $\omega_{i}=\pm\sqrt{\lambda_{i}}$
with $\lambda_{i}$ being the eigenvalues of the Jacobian matrix $\B J$:
\begin{equation}
	\label{deflam}
	\lambda_{1,2}=1\mp\delta\sqrt{1-\nu^{2}},
\end{equation}
where $\nu=\frac{\eta}{\delta}$. Clearly the system develops a complex
pair of eigenvalues for $\eta>1$.  We now take $\epsilon=1-\nu=1-\frac{\eta}{\delta}$ and, in order
for the system to be critical, assume that $\epsilon\ll1$ and positive. Using Eq.~(\ref{deflam}), the associated
frequencies are to leading order
\begin{equation}
	\omega_{1,2} \approx 1 \mp \delta \sqrt{\epsilon/2}\ .
\end{equation}

The eigenvectors $\tilde{\B v}_{1,2}=\left(X,Y\right)$
are obtained as
\begin{equation}
	\tilde {\B v}_{1,2}=\begin{pmatrix}1\pm\sqrt{1-\nu^{2}}\\
		\nu
	\end{pmatrix}.\label{eq:eigvec}
\end{equation}
In the limit $\nu\to 1$ the two critical eigenvectors coincide and become (1,1). 
The two eigenvectors Eq.~(\ref{eq:eigvec})
then become, after normalizing such that $\B v_{1,2} \equiv \tilde {\B v}_{1,2}/\left|\tilde {\B v}_{1,2}\right|=1+O\left(\sqrt{\epsilon}\right)$,
\begin{equation}
	\B v_{1,2}=\frac{1}{\sqrt{2}}\begin{pmatrix}1\pm\sqrt{\epsilon/2}\\
		1\mp\sqrt{\epsilon/2}
	\end{pmatrix}.\label{eq:eigvecnorm}
\end{equation}
 From Eq.~(\ref{eq:eigvecnorm}) we can observe that near the critical
point, the two eigenvectors become parallel. As a result, an initial
condition $(x_0,y_0)$ orthogonal to $\boldsymbol{v^{*}}=\left(1,1\right)/\sqrt{2}$
will result in oscillations whose amplitude diverges as $\epsilon^{-1/2}$.

The effect of an external oscillatory perturbation with an arbitrary
frequency $\omega$ is then studied by adding such a term to the normal form.
\begin{equation}
	\partial_{t}^{2}
	\begin{pmatrix}x\\
		y
	\end{pmatrix}=
	-\begin{pmatrix}1-\delta & \eta\\
		-\eta & 1+\delta
	\end{pmatrix}\begin{pmatrix}x\\
		y
	\end{pmatrix}+\boldsymbol{f}\cos\omega t\label{eq:forcing}
\end{equation}
with $\boldsymbol{f}=F\left(1,-1\right)/\sqrt{2}$ chosen so that
$\boldsymbol{f}\bot\boldsymbol{v^{*}}$. 

The divergence is seen in the homogenous solution to 
(\ref{eq:forcing}) with $\boldsymbol{x}_{{\rm hom}}\left(0\right)=\left(1,-1\right)/\sqrt{2}$.
In Ref.~\cite{20CGPS}, it was shown that
\begin{equation}
	\begin{aligned}x_{{\rm hom},\bot}
		& \approx\frac{1}{\sqrt{\epsilon}}\begin{pmatrix}1\\
			1
		\end{pmatrix}\sin\left(t\right)\sin\left(\frac{\delta \epsilon^{1/2}}{\sqrt{2}}t\right)\\
		& +\frac{1}{\sqrt{2}}\begin{pmatrix}1\\
			-1
		\end{pmatrix}\cos\left(t\right)\cos\left(\frac{\delta \epsilon^{1/2}}{\sqrt{2}}t\right).
	\end{aligned}
	\label{eq:xhom}
\end{equation}
The important conclusions to be drawn for the present study is that the
amplitude $A$ of the homogeneous solution goes as
\begin{equation}
	A\sim\epsilon^{-1/2} \ . \label{ampeps}
\end{equation}

The reader should note that when this normal form is embedded in a {\em nonlinear} system,
the increase in oscillations can easily ignite the nonlinear terms and drive the
system further from equilibrium. Thus one may not see the linear blow-up in its entirety
because nonlinearities will become dominant. An example is shown in the next subsection.

\subsubsection{The effect of dynamical friction}

\begin{figure}
	\includegraphics[scale=0.55]{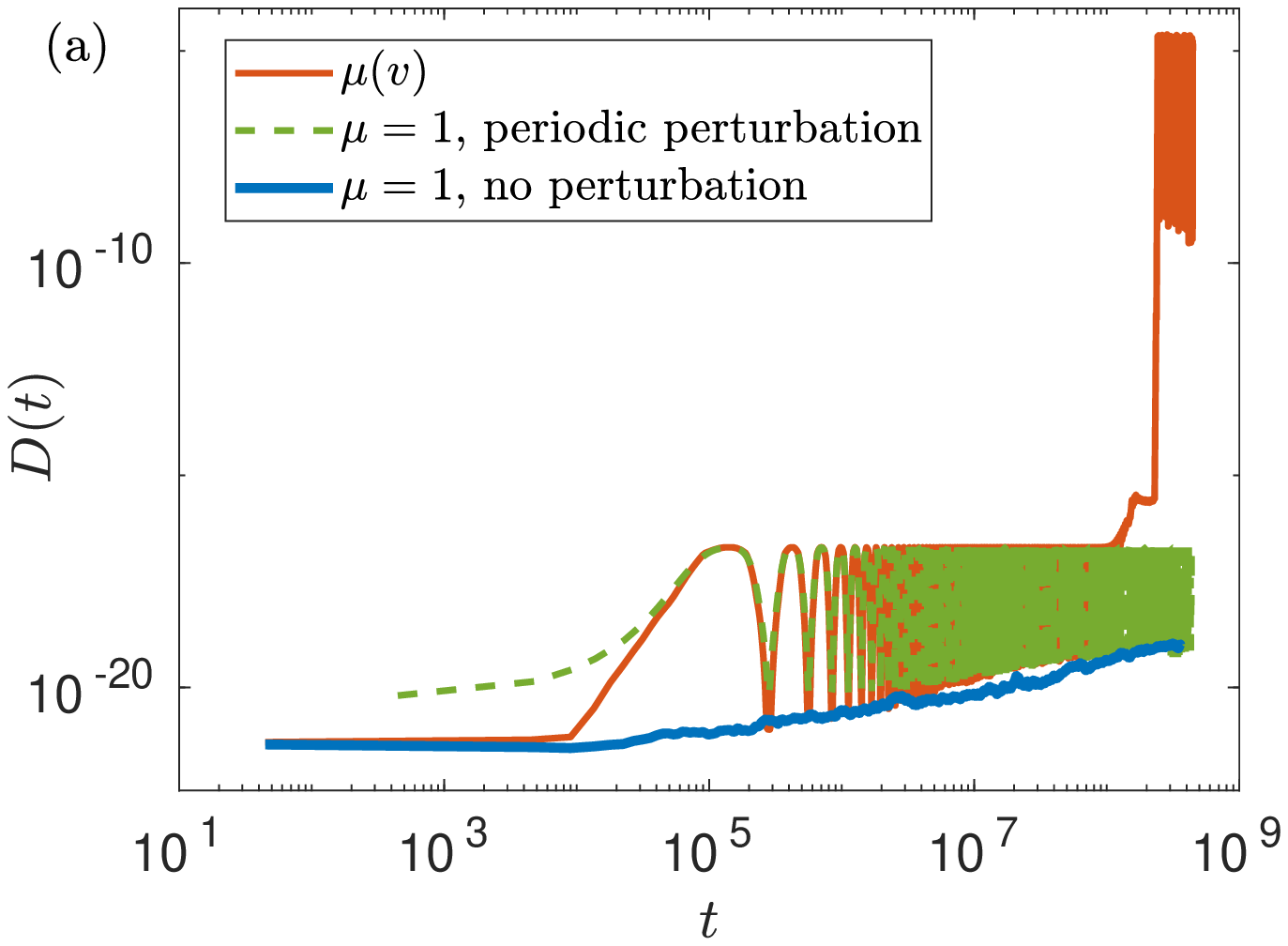}
		\includegraphics[scale=0.55]{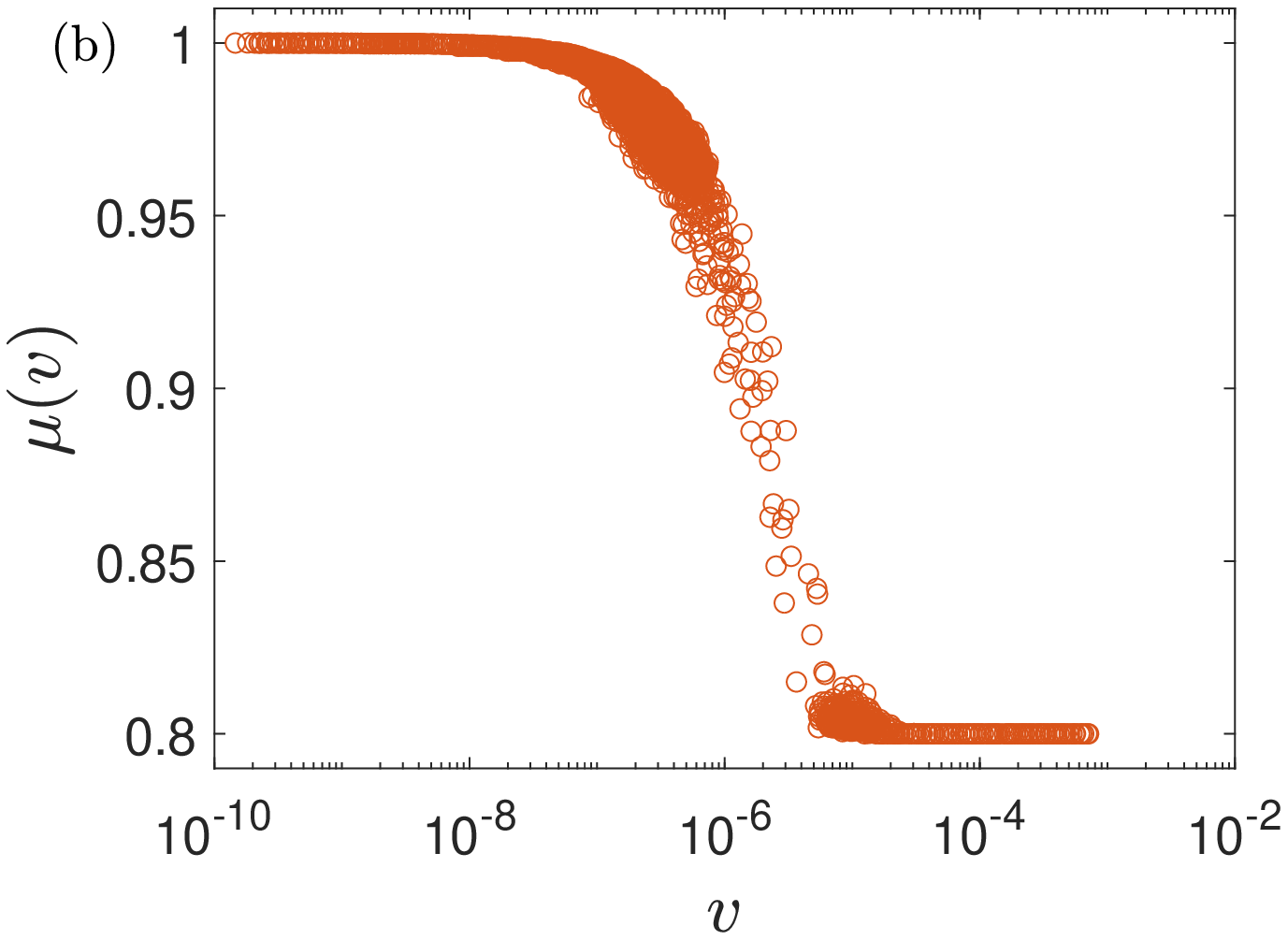}
	\caption{(a) Dynamic response $D(t)$ for the stable system with $F_x=0.0091$ and $\mu_0=1$. If no perturbation is applied (blue line), the system is stable. A minute off-resonance perturbation leads to oscillatory motion (green line), but the system still does not erupt.  Turning on the velocity dependence of $\mu(v)$ (red line) enhances the instability resulting in a giant increase of $D(t)$. (b) The actual $\mu(v)$ dependence. Note the logarithmic scale of the velocity $v$ that changes over more than 7 orders of magnitude.}
	\label{instab} .
\end{figure}

To show that  the dynamic friction can trigger the instability in otherwise stable system, we choose a value of $F_x=0.0091$ that corresponds to a linearly stable configuration for $\mu_0=1$, see inset in Fig. \ref{fdep}. 
Indeed, direct simulations without any damping confirm the stability, cf. the blue line in Fig.~\ref{instab}. The mean-square displacement remains at the level of $10^{-20}$.  

Next, we add a periodic perturbation $f(t)=a \cos{\omega t}$ with a small amplitude $a=10^{-12}$ and a frequency far from resonance $\omega=0.01$. As said above, near onset the typical natural frequency of the oscillatory instability is of the order
of $10^{-5}$. The perturbation has a much slower time scale, but it triggers periodic oscillations in the MSD, with  magnitude still not exceeding $10^{-17}$, cf. the green line in Fig.~\ref{instab}.

But when we allow the friction coefficient to depend on the mean velocity of the upper layer, the oscillatory instability is greatly enhanced, with the MSD increasing by more than 12 orders of magnitude as is shown with the red line in Fig.~\ref{instab}.

In panel (b) of Fig.~\ref{instab}, we present the actual simulated
values of $\mu(v)$. Note that the velocity increases by seven orders
of magnitude. 

\section{The effects of viscous damping}
\label{damping}

In assessing the applicability of the present model to real earthquake dynamics, one needs to state that so far we did not take into account any mechanism of damping. Damping can arise due to a variety of causes, not the least being the presence of water mixed with the gouge in the fault.  It is thus important to investigate the consequences of damping on the normal form that we studied above and on the giant sensitivity to small perturbations.

Let us then consider the case in which a damping force
\begin{equation}
	\boldsymbol{f}_{{\rm damp}}=-\frac{1}{\tau}\partial_{t}\begin{pmatrix}x\\
		y
	\end{pmatrix}
\end{equation}
is added to equation (\ref{eq:forcing}).

For the homogeneous solution (\ref{eq:xhom}), we note that $\boldsymbol{v}_{1,2}$
from equation (\ref{eq:eigvecnorm}) are still the eigenvectors
of the system near criticality. A homogeneous solution of the form
$\left(X,Y\right)e^{i\omega t}$ will now solve
\begin{equation}
	\left[-\begin{pmatrix}\omega^{2}-i\omega/\tau & 0\\
		0 & \omega^{2}-i\omega/\tau
	\end{pmatrix}+\begin{pmatrix}1-\delta & \eta\\
		-\eta & 1+\delta
	\end{pmatrix}\right]\begin{pmatrix}X\\
		Y
	\end{pmatrix}=0 \ ,
\end{equation}
so that $\boldsymbol{v}_{1,2}$ is still obtained from (\ref{eq:eigvecnorm}).
In the case of a small damping ($1/\tau \ll 1$), Eq.~(\ref{eq:xhom})
becomes (cf. Ref~\cite{20CGPS}) 
\begin{equation}
	\begin{aligned}x_{{\rm hom}} & \approx\frac{1}{\sqrt{\epsilon}}\begin{pmatrix}1\\
			1
		\end{pmatrix}\sin\left(t\right)\sin\left(\frac{\delta \epsilon^{1/2}}{\sqrt{2}}t\right)e^{-\frac{t}{\tau}}\\
		& +\frac{1}{\sqrt{2}}\begin{pmatrix}1\\
			-1
		\end{pmatrix}\cos\left(t\right)\cos\left(\frac{\delta \epsilon^{1/2}}{\sqrt{2}}t\right)e^{-\frac{t}{\tau}} \ .
	\end{aligned}
\end{equation}
The maximal amplitude is obtained as
\begin{equation}
	A\sim\max_{t}\left[\frac{1}{\sqrt{\epsilon}}\sin\left(t/\tau_d\right)e^{-\frac{t}{\tau}}\right] \ ,
\end{equation}
with $\tau_d \sim \frac{1}{\delta\sqrt{\epsilon}}$.

At this point it becomes clear that there is a competition between the damping
time scale $\tau$ and the closeness to criticality as measured by $\tau_d$. 
If $\tau_d\ll \tau$, the damping is not strongly effective
and we expect to recover Eq.~(\ref{ampeps}). On the other hand, in the opposite
limit, $\tau_d\gg \tau$, we expect to lose much of the effect of the giant sensitivity,
and seek an explanation for remote triggering, if indeed occurring in this limit, with
another mechanism.

\section{summary and conclusions}
\label{summary}

In this paper we offered a possible mechanism for remote triggering, paying attention to the requirements described in the introduction: (i) linear stability under stress, (ii) giant sensitivity to small
perturbations, (iii) the small perturbations can have an arbitrary frequency, and (iv) genericity. All these requirements are satisfied by the proposed mechanics which rests on the breaking of Hamiltonian symmetry due to friction. The weak point of our scenario is that strong dissipative effects can destroy the giant sensitivity. It therefore remains to examine actual cases of remote triggering to assess whether in real units the magnitude and nature of the dissipative effects leave the present proposed mechanism relevant to earthquakes in the field. This certainly appears as a worthwhile research objective.   

\appendix

\section{Tangential displacements}
\label{tang}

The normal force laws were described in detail in the main text. Here we
spell out the calculation of the tangential forces. 
The tangential displacements can be divided into two parts: 
(1) disk-disk interaction and (2) disk-wall interaction

{\bf Disk-disk pair}: The tangential velocity which is the derivative of the tangential displacement {\bf t}$_{ij}$ of a pair 
of interacting disks $i$ and $j$ is given by:
\begin{equation}\label{eq11}
	{\bf v}_{ij}^{t} = \frac{\textrm{d}{\bf t}_{ij}}{\textrm{d}t} = {\bf v}_{ij} - {\bf v}^{n}_{ij} + 
	\hat{r}_{ij} \times (R_{i}\boldsymbol{\omega}_{i} + R_{j}\boldsymbol{\omega}_{j})
\end{equation}
where, ${\bf v}_{ij} = {\bf v}_{i} - {\bf v}_{j}$ is the relative velocity of the pair-$i$ and $j$. 
${\bf v}_{ij}^{n}$ is the projection of the ${\bf v}_{ij}$ along the direction $\hat{r}_{ij}$.
$\boldsymbol{\omega}_{i}$ and $\boldsymbol{\omega}_{j}$ are the angular velocities of the disks $i$ 
and $j$ respectively given by $\boldsymbol{\omega}_{ij} = \frac{\textrm{d}\boldsymbol{\theta}_{ij}}{\textrm{d}t}$,
here, $\textrm{d}\boldsymbol{\theta}_{ij}$ is the angular displacement of disk $i$ with respect to disk $j$.
The above equation can be written in differential form as follow:
\begin{equation}\label{eq13}
	\textrm{d}{\bf t}_{ij} = \textrm{d}{\bf r}_{ij} - \textrm{d}{\bf r}^{n}_{ij} + 
	\hat{r}_{ij} \times (R_i\textrm{d}\boldsymbol{\theta}_{i} + R_j\textrm{d}\boldsymbol{\theta}_{j})
\end{equation}
For a quasi-two-dimensional system as is in our studies, $\boldsymbol{\omega}_{i}$ and 
$\boldsymbol{\theta}_{i}$ will have only one component in $\hat{z}$-direction perpendicular to the 
$xy$-plane, therefore, 
$\hat{r}_{ij} \times R_i\textrm{d}\boldsymbol{\theta}_{i} = 
R_i\textrm{d}\theta_{i}(y_{ij}\hat{x} - x_{ij}\hat{y})/r_{ij}$. Hence, Eq.(\ref{eq13}) can be written in 
tensorial form as follow:
\begin{equation}\label{eq14}
	\textrm{d}t^{\alpha}_{ij} = \textrm{d}r^{\alpha}_{ij} - (\textrm{d}{\bf r}_{ij}\cdot\hat{r}_{ij})\frac{r^{\alpha}_{ij}}{r_{ij}} 
	+ (-1)^{\alpha}(R_i\textrm{d}\theta_{i}+R_j\textrm{d}\theta_{j})\frac{r^{\beta}_{ij}}{r_{ij}}  \ ,
\end{equation}
where $\alpha$ and $\beta$ correspond to $x$ and $y$ components, respectively.

\section{Dynamic friction}
\label{dynfric}

When a system of blocks or disks subjected to a normal force $\B F_y$, is pulled with a  constant velocity (or a constant force) in the $x$-direction, at first no sliding occur. Once the pulling velocity become larger than some threshold $v_0$, the system starts to slide.  At small velocities, a non-steady stick-slip motion is taking place. At larger velocities, the sliding become steady. The friction coefficient $\mu(v)$ depends on the pulling velocity and decreases from the static value $\mu_0$  to the value corresponding to the steady sliding. Both friction coefficients $\mu_0$ and $\mu(v)$ are not fully determined by the material constants, they depend also on the protocol, e.g. the time that the surfaces spent in static contact, the pulling velocity and, in the experiments involving a change of the pulling velocity, the ratio of the initial and final  velocities. 
This variability is attributed to the changes in the structure of the real contact area between two surfaces.
\begin{figure}
	\includegraphics[scale=0.6]{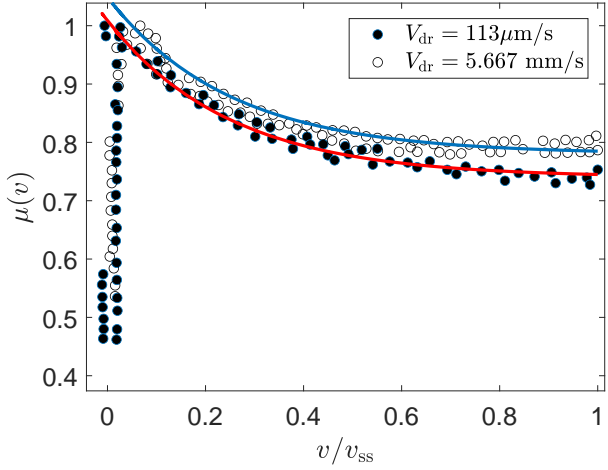}
	\caption{The velocity dependence of the friction coefficient for experiments in Ref.~\cite{98NKBG}. Open and filled circles denote the data of measurement for two driving velocities as indicated in the figure. The solid lines are fits with the functional form Eq.\eqref{muv}. The fit parameters $B=0.22, w=4, \tilde v_0=0.05$. The steady-slide friction coefficeint $\mu_{\rm ss}=0.74$ for filled circles and $\mu_{\rm ss}=0.78$  for open circles. The experimental data are reproduced from  Fig. 13 in Ref~\cite{98NKBG}. }
	\label{muexp}
\end{figure}

In natural conditions, the surfaces in contact are formed by grains of various sizes. The sliding motion involve thousands of grains and both the grain displacement and slide velocities may vary by many order of magnitude \cite{97NKG,98NKBG,04DiToro,Dieterich78,AharonovScholz18}.
A good idea of the velocity-dependence of the friction coefficient at small driving velocities is provided by experiments with large spherical glass particles \cite{98NKBG}. The driving velocity is applied to a cover glass, while the role of the normal force is played by the gravity force $F_g$. The friction force $F_f$ is uniquely defined by the driving velocity and the friction coefficient (or the normalized instantaneous frictional force) $\mu(t)=F_f/F_g$, measured together with the deflection rate of the cover glass. At slow driving velocities, the motion of the cover glass is unsteady, with fast slip events followed by almost stationary "stick" periods.

Although the particular dependence of the instantaneous friction coefficient on the instantaneous velocity is influenced by material parameters and driving velocity, the shape $\mu(v)$ appears almost universal, when plotted in a dimensionless form $\mu(v)/\mu_0$ vs $v/v_{\rm ss}$, where $\mu_0$ is the friction coefficient measured before any sliding occur, and  $v_{\rm ss}$ is the largest  velocity  measured during slip event. This shape, given by Eq.\eqref{muv}, fits well with almost identical fitting parameters the results measured for two very different driving velocities, as illustrated in Fig. \ref{muexp}.

\section{Structure of Jacobian}\label{jacobian}

The evaluation of the $\B J$ operator is described in two section here. We first provide the calculations for 
the disk-disk pair which is followed by the disk-wall interaction. 

{\bf Disk-disk pair}: 
Derivative of the tangential force (Eq.(\ref{eq14})) with respect to $r^{\alpha}_{i}$:
\begin{widetext}
	\begin{equation}\label{eq26}
		\begin{array}{l}
			\pdv{{\textnormal{F}^{(t)}_{ij}}^{\beta}}{r^{\alpha}_{j}} = 
			-k_{t} \pdv{}{r^{\alpha}_{i}}\Bigg[\delta_{ij}^{1/2}\Bigg(t^{\beta}_{ij} + \tilde{t}t^{\beta}_{ij} - 
			\tilde{t}^{2}t^{\beta}_{ij}\Bigg)\Bigg]\\
			= \frac{1}{2} \delta^{-1}_{ij} \frac{r_{ij}^{\alpha}}{r_{ij}} {\textnormal{F}_{ij}^{(t)}}^{\beta} 
			-\,\,k_{t}\delta^{1/2}_{ij}\Bigg[(1 + \tilde{t} - \tilde{t}^{2})\pdv{t^{\beta}_{ij}}{r^{\alpha}_{i}}
			\,\,+\,\,(\tilde{t}^{\beta} - 2\tilde{t}\tilde{t}^{\beta})\pdv{t_{ij}}{r^{\alpha}_{i}} 
			\,\,+\,\,(-\tilde{t}\tilde{t}^{\beta} + 2\tilde{t}^{2}\tilde{t}^{\beta})\pdv{t^{\star}_{ij}}{r^{\alpha}_{i}}\Bigg]
		\end{array}
	\end{equation}
\end{widetext}
Here, $\tilde{t} = \frac{t_{ij}}{t^{\star}_{ij}}$ and  
$\tilde{t}^{\beta} = \frac{t_{ij}^{\beta}}{t^{\star}_{ij}}$. 
Expressions for all the three partial differentiation in (Eq.(\ref{eq26})) are shown in \cite{20CGPS}.

Similarly, the derivative of tangential force with respect to $\theta_{j}$ (using the same notation as
above) can be found as:
\begin{equation}\label{eq27}
	\pdv{{\textnormal{F}^{(t)}_{ij}}^{\beta}}{\theta_{j}} = 
	-k_{t} \delta^{\frac{1}{2}}\Bigg[(1 + \tilde{t} - \tilde{t}^{2})\pdv{t^{\beta}_{ij}}{\theta_{j}} 
	+ (\tilde{t}^{\beta} - 2\tilde{t}\tilde{t}^{\beta})\pdv{t_{ij}}{\theta_{j}}\Bigg]
\end{equation}
From the above two equations, it is then understood that if $r_{ij}$ and $t_{ij}$ are known, the 
differential equations can be solved \cite{20CGPS}. When $\tilde{t}^{\beta}$ is negligible for all $\beta$, then 
$\tilde{t} \approx 0$, which then translates to 
$\pdv{\textnormal{F}^{(t)}_{ij}}{\theta_{j}} =  -k_{t} \delta^{\frac{1}{2}}\pdv{t^{\beta}_{ij}}{\theta_{j}} = -
(-1)^{\beta}\,R_i\delta^{\frac{1}{2}}_{ij}\,\frac{r_{ij}^{\alpha}}{r_{ij}}$, 
with $\alpha \ne \beta$, implying that even in the case of zero tangential displacement [hence, zero
tangential force], the above derivative can be finite.

The derivative of the normal force with respect to $r_{j}^{\alpha}$ follows:
\begin{equation}\label{eq28}
	\begin{array} {lcl}
		\pdv{{\textnormal{F}^{(n)}_{ij}}^{\beta}}{r^{\alpha}_{j}} & = & 
		k_{n}\pdv{}{r^{\alpha}_{j}}\Bigg[\delta^{\frac{3}{2}}_{ij} \frac{r^{\alpha}_{ij}}{r_{ij}}\Bigg] \\
		& = &  k_{n}\delta^{\frac{1}{2}}_{ij}\Bigg[-\Delta_{\alpha\beta}\frac{\delta_{ij}}{r_{ij}}
		+ \frac{3}{2} \frac{r^{\alpha}_{ij}r^{\beta}_{ij}}{r^{2}_{ij}}  
		+ \Big(\frac{\delta_{ij}}{r_{ij}}\Big)\frac{r^{\alpha}_{ij}r^{\beta}_{ij}}{r^{2}_{ij}}\Bigg] 
	\end{array}
\end{equation}
where $\Delta_{\alpha\beta}$ is the Kronecker delta. 
Therefore, the derivative of the total force is written as follow:
\begin{equation}\label{eq29}
	\pdv{\textnormal{F}^{\beta}_{ij}}{r^{\alpha}_{j}} = 
	\pdv{\textnormal{F}^{(n)\beta}_{ij}}{r^{\alpha}_{j}} + \pdv{\textnormal{F}^{(t)\beta}_{ij}}{r^{\alpha}_{j}}
\end{equation}
\begin{equation}\label{eq30}
	\pdv{\textnormal{F}^{\beta}_{ij}}{\theta_{j}} = \pdv{\textnormal{F}^{(t)\beta}_{ij}}{\theta_{j}}
\end{equation}

The torque of a disk-$j$ due to tangential force 
$\textnormal{\textbf{F}}^{(t)}_{ij}$ is $\textbf{T}_{j} = 
-R_j(\hat{r}_{ij} \times \textbf{\textnormal{F}}^{(t)}_{ij}) \equiv R_j\tilde{\textnormal{\textbf{T}}}_{ij}$.
In two-dimensional systems, $\tilde{\textnormal{\textbf{T}}}_{ij}$ has only $z$-component:
\begin{equation}\label{eq31}
	\tilde{\textnormal{T}}^{z}_{ij} = -\Bigg[\Bigg(\frac{x_{ij}}{r_{ij}}\Bigg){\textnormal{F}^{(t)}_{ij}}^{y}
	- \Bigg(\frac{y_{ij}}{r_{ij}}\Bigg){\textnormal{F}^{(t)}_{ij}}^{x} \Bigg]
\end{equation}
Therefore, the derivative of $\tilde{\textnormal{T}}^{z}_{ij}$ with respect to translational coordinates 
$r^{\alpha}_{i}$ then becomes:
\begin{widetext}
	\begin{equation}\label{eq32}
		\pdv{\tilde{\textnormal{T}}^{z}_{ij}}{r^{\alpha}_{j}} = 
		\Bigg(\frac{\Delta_{\alpha x}}{r_{ij}} - \frac{x_{ij}x^{\alpha}_{ij}}{r^{3}_{ij}}\Bigg){\textnormal{F}^{(t)}_{ij}}^{y}
		- \Bigg( \frac{x_{ij}}{r_{ij}}\Bigg)\pdv{{\textnormal{F}^{(t)}_{ij}}^{y}}{r^{\alpha}_{j}}
		- \Bigg(\frac{\Delta_{\alpha y}}{r_{ij}} - \frac{y_{ij}x^{\alpha}_{ij}}{r^{3}_{ij}}\Bigg){\textnormal{F}^{(t)}_{ij}}^{x}
		+ \Bigg( \frac{y_{ij}}{r_{ij}}\Bigg)\pdv{{\textnormal{F}^{(t)}_{ij}}^{x}}{r^{\alpha}_{j}}
	\end{equation}
\end{widetext}
where $\Delta_{\alpha x}$ and, $\Delta_{\alpha y}$ is the Kronecker delta, i.e. 
$\Delta_{xx} = \Delta_{yy} = 1$ and zero otherwise. 
The derivative of the torque with respect to the $\theta_{i}$ would be:
\begin{equation}\label{eq33}
	\pdv{\tilde{\textnormal{T}}^{z}_{ij}}{\theta_{j}} = -\Bigg[
	\Bigg(\frac{x_{ij}}{r_{ij}}\Bigg)\pdv{\textnormal{F}^{y}_{ij}}{\theta_{j}}
	-\Bigg(\frac{y_{ij}}{r_{ij}}\Bigg)\pdv{\textnormal{F}^{x}_{ij}}{\theta_{j}}
	\Bigg]
\end{equation}
The above two differential equations can be solved using Eq.(\ref{eq26}), and Eq.(\ref{eq27}). If the 
tangential displacement $t^{\beta}_{ij}$ is negligible compared to the threshold $t^{\star}_{ij}$, 
i.e., $\tilde{t}^{\beta}_{ij}  \approx 0$ for all $\beta$, this provides us with $\tilde{t} \approx 0$.
Therefore, $\pdv{\tilde{\textnormal{T}}^{z}_{ij}}{\theta_{i}} = k_{t}R_i\delta^{\frac{1}{2}}_{ij}$, which
means that even in the case of negligible/zero tangential displacement hence zero tangential force, 
the derivative of the torque is non-zero. 

{\bf Disk-wall interaction}: 
Following Eq.(\ref{eq26}) for the disk-disk pair, we can write the derivative of the tangential 
force with respect $r^{\alpha}_{i}$ for the disk-wall pair as:
\begin{widetext}
	\begin{equation}\label{eq34}
		\begin{array}{l}
			\pdv{{\textnormal{F}^{(t)}_{iw}}^{\beta}}{r^{\alpha}_{i}} = 
			-k_{t} \pdv{}{r^{\alpha}_{i}}\Bigg[\delta_{iw}^{1/2}\Bigg(t^{\beta}_{iw} + \tilde{t}_{w}t^{\beta}_{iw} - 
			\tilde{t}_{w}^{2}t^{\beta}_{iw}\Bigg)\Bigg]\\
			= -\frac{1}{2} \delta^{-1}_{iw} \frac{r_{iw}^{\alpha}}{r_{iw}} {\textnormal{F}_{iw}^{(t)}}^{\beta} 
			-\,\,k_{t}\delta^{1/2}_{iw}\Bigg[(1 + \tilde{t}_{w} - \tilde{t}_{w}^{2})\pdv{t^{\beta}_{iw}}{r^{\alpha}_{i}}
			\,\,+\,\,(\tilde{t}_{w}^{\beta} - 2\tilde{t}_{w}\tilde{t}_{w}^{\beta})\pdv{t_{iw}}{r^{\alpha}_{i}} 
			\,\,+\,\,(-\tilde{t}_{w}\tilde{t}_{w}^{\beta} + 2\tilde{t}_{w}^{2}\tilde{t}_{w}^{\beta})\pdv{t^{\star}_{iw}}{r^{\alpha}_{i}}\Bigg]
		\end{array}
	\end{equation}
\end{widetext}
Here, $\tilde{t}_{w} = \frac{t_{iw}}{t^{\star}_{iw}}$ and  
$\tilde{t}^{\beta}_{w} = \frac{t_{iw}^{\beta}}{t^{\star}_{iw}}$.

Similarly, the derivative of tangential force with respect to $\theta_{i}$ (using the same notation as
above) can be found as:
\begin{equation}\label{eq35}
	\pdv{\textnormal{F}^{(t)}_{iw}}{\theta_{i}} = 
	-k_{t} \delta^{\frac{1}{2}}\Bigg[(1 + \tilde{t}_{w} - \tilde{t}_{w}^{2})\pdv{t^{\beta}_{iw}}{\theta_{i}} 
	+ (\tilde{t}_{w}^{\beta} - 2\tilde{t}_{w}\tilde{t}_{w}^{\beta})\pdv{t_{iw}}{\theta_{i}}\Bigg]
\end{equation}

The derivative of the normal Hertzian force would be:
\begin{equation}\label{eq36}
	\begin{array} {lcl}
		\pdv{{\textnormal{F}^{(n)}_{iw}}^{\beta}}{r^{\alpha}_{i}} & = & 
		k_{n}\pdv{}{r^{\alpha}_{i}}\Bigg[\delta^{\frac{3}{2}}_{iw} \frac{r^{\alpha}_{iw}}{r_{iw}}\Bigg] \\
		& = &  k_{n}\delta^{\frac{1}{2}}_{iw}\Bigg[\Delta_{\alpha\beta}\frac{\delta_{iw}}{r_{iw}}
		- \frac{3}{2} \frac{r^{\alpha}_{iw}r^{\beta}_{ij}}{r^{2}_{ij}} - 
		\Big(\frac{\delta_{iw}}{r_{iw}}\Big)\frac{r^{\alpha}_{iw}r^{\beta}_{iw}}{r^{2}_{iw}}\Bigg] 
	\end{array}
\end{equation}
where $\Delta_{\alpha\beta}$ is the Kronecker delta. In the above equation, one should note that
if $\textnormal{F}^{nx}_{iw} = 0$ (which is true in the current studies) 
then $\pdv{{\textnormal{F}^{n}_{iw}}^{\beta}}{r^{\alpha}_{i}} = 0$

Therefore, the derivative of the total force is written as follow:
\begin{equation}\label{eq37}
	\pdv{\textnormal{F}^{\beta}_{iw}}{r^{\alpha}_{i}} = 
	\pdv{\textnormal{F}^{(n)\beta}_{iw}}{r^{\alpha}_{i}} + \pdv{\textnormal{F}^{(t)\beta}_{iw}}{r^{\alpha}_{i}}
\end{equation}
\begin{equation}\label{eq38}
	\pdv{\textnormal{F}^{\beta}_{iw}}{\theta_{i}} = \pdv{\textnormal{F}^{(t)\beta}_{iw}}{\theta_{i}}
\end{equation}

The torque of a disk-$i$ due to tangential force 
$\textnormal{\textbf{F}}^{(t)}_{iw}$ is $\textbf{T}_{i} = 
-R_i(\hat{r}_{iw} \times \textbf{\textnormal{F}}^{t}_{iw}) \equiv R_i\tilde{\textnormal{\textbf{T}}}_{iw}$.
In two-dimensional systems, $\tilde{\textnormal{\textbf{T}}}_{iw}$ has only $z$-component:
\begin{equation}\label{eq39}
	\tilde{\textnormal{T}}^{z}_{iw} = -\Bigg[\Bigg(\frac{x_{iw}}{r_{iw}}\Bigg){\textnormal{F}^{(t)}_{iw}}^{y}
	- \Bigg(\frac{y_{iw}}{r_{iw}}\Bigg){\textnormal{F}^{(t)}_{iw}}^{x} \Bigg]
\end{equation}
Therefore, the derivative of $\tilde{\textnormal{T}}^{z}_{iw}$ with respect to translational coordinates 
$r^{\alpha}_{i}$ then becomes:
\begin{widetext}
	\begin{equation}\label{eq40}
		\pdv{\tilde{\textnormal{T}}^{z}_{iw}}{r^{\alpha}_{i}} = 
		- \Bigg(\frac{\Delta_{\alpha x}}{r_{iw}} - \frac{x_{iw}x^{\alpha}_{iw}}{r^{3}_{iw}}\Bigg){\textnormal{F}^{t}_{iw}}^{y}
		- \Bigg( \frac{x_{iw}}{r_{iw}}\Bigg)\pdv{{\textnormal{F}^{t}_{iw}}^{y}}{r^{\alpha}_{iw}}
		+ \Bigg(\frac{\Delta_{\alpha y}}{r_{iw}} - \frac{y_{iw}x^{\alpha}_{iw}}{r^{3}_{iw}}\Bigg){\textnormal{F}^{t}_{iw}}^{x}
		+ \Bigg( \frac{y_{iw}}{r_{iw}}\Bigg)\pdv{{\textnormal{F}^{t}_{iw}}^{x}}{r^{\alpha}_{iw}}
	\end{equation}
\end{widetext}
where $\Delta_{\alpha x}$ and, $\Delta_{\alpha y}$ are the Kronecker delta, i.e. 
$\Delta_{xx} = \Delta_{yy} = 1$ and zero otherwise. 
The derivative of the torque with respect to the $\theta_{i}$ would be:
\begin{equation}\label{eq41}
	\pdv{\tilde{\textnormal{T}}^{z}_{iw}}{\theta_{i}} = -\Bigg[
	\Bigg(\frac{x_{iw}}{r_{iw}}\Bigg)\pdv{\textnormal{F}^{y}_{iw}}{\theta_{i}}
	-\Bigg(\frac{y_{iw}}{r_{iw}}\Bigg)\pdv{\textnormal{F}^{x}_{iw}}{\theta_{i}}
	\Bigg]
\end{equation}

{\bf Expressions for the different parts of the Jacobian}:
The dimension of Jacobian operator $\B J$ is force over length. To be consistent with the dimension, we 
redefine the torque T and rotational coordinate $\theta_{i}$ as:
\begin{equation}\label{eq42}
	\tilde{\textnormal{T}}_{i} = \frac{\textnormal{T}_{i}}{R_i}  
	\quad \textnormal{and} \quad
	\tilde{\theta_{i}} = R_i\theta_{i}
\end{equation}
In addition, the dynamical matrix has a contribution from the moment of inertia 
$\textnormal{I}_{i} = \textnormal{I}_{0}m_{i}R_i^{2}$ since 
$\Delta \boldsymbol{\omega}_{i} = \frac{\textbf{T}_{i}}{\textnormal{I}_{i}\Delta t}$. In our 
calculation, we assume that mass $m_{i}$ and $\textnormal{I}_{0}$ both are 
one. The remaining contribution of $\textnormal{I}_{i}$ i.e. $R_i^{2}$, is taken care of by
re-scaling the torque and angular displacement as $\tilde{\textnormal{T}}_{i}$ and 
$\tilde{\theta}_{i}$. For $\textnormal{I}_{i} \ne 1$, the contribution of
$\textnormal{I}_{i}$ can be correctly anticipated if we rewrite Eq.(\ref{eq11}) as below:
\begin{equation}\label{eq43}
	\frac{\textnormal{d}\textbf{t}_{ij}}{\textnormal{d}t} = \textbf{v}_{ij} - \textbf{v}_{ij}^{n} 
	+ \frac{1}{\textnormal{I}_{0}}\hat{r}_{ij} \times R_i(\boldsymbol{\omega}_{i} + \boldsymbol{\omega}_{j})
\end{equation}

The Jacobian operator $\B J = \Big|\Big|\pdv{\textbf{F}_{i}}{\textbf{Q}_{j}}\Big|\Big|_{\bf Q_{{\bf 0}}}$
can be divided into four parts, given as: (i) Derivative of the force $F^{\beta}_{i}$ with respect to 
translational coordinates $r^{\alpha}_{j}$ (ii) Derivative of the force $F^{\beta}_{i}$ with respect to 
angular displacements $\theta_{j}$ (iii) Derivative of the torque $T^{\beta}_{i}$ with respect to 
translational coordinates $r^{\alpha}_{j}$ (iv) Derivative of the torque $F^{\beta}_{i}$ with respect 
to angular displacements $\theta_{j}$. 
Total force on disk $i$ is given by:
\begin{equation}\label{eq44}
	\textbf{F}_{i} = \sum_{j = 1, i \ne j}^{N} \textbf{F}_{ij} + \textbf{F}_{iw} + \textbf{F}_{i}^{y} 
	+ \textbf{F}_{i}^{x}
\end{equation}
where,
\begin{enumerate}[label={(\roman*)}]
	\item $\sum_{j}^{N} \textbf{F}_{ij}$ = total force on the disk due to other disk in contact
	\item $\textbf{F}_{iw}$ = total force on the disk due to wall
	\item $\textbf{F}_{i}^{y}$ = constant force on the upper layer disks pushing the disks down 
	(-$\hat{y}$ direction)
	\item $\textbf{F}_{i}^{x}$ = constant force on all the disks of the upper layer, pushing the disks in the horizontal
	(+$\hat{x}$ direction). 
\end{enumerate}
Differentiating both sides with respect to $\bf{r}_{j}$\\
$$
\pdv{\textbf{F}_{i}}{\textbf{r}_{j}} = 
\frac{\partial}{\partial\textbf{r}_{j}}\sum_{j=1,j\ne i}^{N} \textbf{F}_{ij}
+ \frac{\partial}{\partial\textbf{r}_{j}}\textbf{F}_{iw}
$$ 
\begin{equation}\label{eq45}
	\pdv{\textbf{F}_{i}}{\textbf{r}_{j}} = 
	\frac{\partial}{\partial\textbf{r}_{j}}\sum_{j=1,j\ne i}^{N} \textbf{F}_{ij}
	+ \frac{\partial}{\partial\textbf{r}_{j}}\textbf{F}_{iw}
\end{equation}
Using tensor notations, we get
\begin{equation}\label{eq46}
	\pdv{F^{\beta}_{i}}{r^{\alpha}_{j}} = 
	\frac{\partial}{\partial r^{\alpha}_{j}}\sum_{j=1,j\ne i}^{N} F^{\beta}_{ij}
	+ \frac{\partial F^{\beta}_{iw}}{\partial r^{\alpha}_{j}}
\end{equation}
{\bf First part:} Derivative of forces with respect to positions of particles:\\
\begin{equation}\label{eq49}
	J^{\alpha \beta}_{ij} = \frac{\partial F^{\beta}_{ij}}{\partial r^{\alpha}_{j}};\hspace{1cm}\,for\,\, j \ne i
\end{equation}
\begin{equation}\label{eq50}
	\pdv{F^{\beta}_{i}}{r^{\alpha}_{i}} = -\sum_{j=1,j\ne i}^{N}\frac{\partial F^{\beta}_{ij}}{\partial r^{\alpha}_{j}} 
	+ \frac{\partial F^{\beta}_{iw}}{\partial r^{\alpha}_{i}}
	;\hspace{1cm}\,for\,\, j = i
\end{equation}
which then gives us
\begin{equation}\label{eq51}
	\begin{aligned}
		J^{\alpha \beta}_{ij} &= \frac{\partial F^{\beta}_{ij}}{\partial r^{\alpha}_{j}}\\\\
		J^{\alpha \beta}_{ii} &= -\sum_{j=1,j\ne i}^{N}\frac{\partial F^{\beta}_{ij}}{\partial r^{\alpha}_{j}} 
		+ \frac{\partial F^{\beta}_{iw}}{\partial r^{\alpha}_{i}}
	\end{aligned}
\end{equation}
{\bf Second part}: Derivative of forces with respect to rotational coordinates of particles:\\
\begin{equation}\label{eq55}
	J^{\beta}_{ij} = \frac{\partial F^{\beta}_{ij}}{\partial \theta_{j}};\hspace{1cm}\,for\,\, j \ne i
\end{equation}
\begin{equation}\label{eq56}
	\pdv{F^{\beta}_{i}}{\theta_{i}} = \sum_{j=1,j\ne i}^{N}\frac{\partial F^{\beta}_{ij}}{\partial \theta_{j}} 
	+ \frac{\partial F^{\beta}_{iw}}{\partial \theta_{i}}
	;\hspace{1cm}\,for\,\, j = i
\end{equation}
which gives
\begin{equation}\label{eq57}
	\begin{aligned}
		J^{\beta}_{ij} &= \frac{\partial F^{\beta}_{ij}}{\partial \theta_{j}}\\\\
		J^{\beta}_{ii} &= \sum_{j=1,j\ne i}^{N}\frac{\partial F^{\beta}_{ij}}{\partial \theta_{j}} 
		+ \frac{\partial F^{\beta}_{iw}}{\partial \theta_{i}}
	\end{aligned}
\end{equation}
{\bf Third part}
Derivative of torques with respect to positions of particles:\\
\begin{equation}\label{eq60}
	J^{\alpha}_{ij} = \frac{\partial T^{z}_{ij}}{\partial r^{\alpha}_{j}};\hspace{1cm}\,for\,\, j \ne i
\end{equation}
\begin{equation}\label{eq61}
	\pdv{T^{z}_{i}}{r^{\alpha}_{i}} = -\sum_{j=1,j\ne i}^{N}\frac{\partial T^{z}_{ij}}{\partial r^{\alpha}_{j}} 
	+ \frac{\partial T^{z}_{iw}}{\partial r^{\alpha}_{i}}
	;\hspace{1cm}\,for\,\, j = i
\end{equation}
which gives
\begin{equation}\label{eq62}
	\begin{aligned}
		J^{\alpha}_{ij} &= \frac{\partial T^{z}_{ij}}{\partial r^{\alpha}_{j}}\\\\
		J^{\alpha}_{ii} &=  -\sum_{j=1,j\ne i}^{N}\frac{\partial T^{z}_{ij}}{\partial r^{\alpha}_{j}} 
		+ \frac{\partial T^{z}_{iw}}{\partial r^{\alpha}_{i}}
	\end{aligned}
\end{equation}
{\bf Fourth part}
Derivative of torques with respect to rotational coordinates of particles\\
\begin{equation}\label{eq65}
	J_{ij} = \frac{\partial T^{z}_{ij}}{\partial \theta_{j}};\hspace{1cm}\,for\,\, j \ne i
\end{equation}
\begin{equation}\label{eq66}
	\pdv{T^{z}_{i}}{\theta_{i}} = \sum_{j=1,j\ne i}^{N}\frac{\partial T^{z}_{ij}}{\partial \theta_{j}} 
	+ \frac{\partial T^{z}_{iw}}{\partial \theta_{i}}
	;\hspace{1cm}\,for\,\, j = i
\end{equation}
which gives
\begin{equation}\label{eq67}
	\begin{aligned}
		J_{ij} &= \frac{\partial T^{z}_{ij}}{\partial \theta_{j}}\\\\
		J_{ii} &=  \sum_{j=1,j\ne i}^{N}\frac{\partial T^{z}_{ij}}{\partial \theta_{j}} 
		+ \frac{\partial T^{z}_{iw}}{\partial \theta_{i}}
	\end{aligned}
\end{equation}

{\bf An example of the Jacobian matrix for a 2 disks system}:
For each disk the total degrees of freedom are 3, i.e. two  in $x$ and $y$ and the third one is the 
$\theta$ rotation. Hence, the dimension of the Jacobian matrix is $(d + 1)N\times(d + 1)N$. Here we 
show the arrangement of the elements of the Jacobian matrix for a two disks system. 
\begin{center}
	\begin{math}
		J = 
		\begin{bmatrix}
			\frac{\partial F_{1}^{x}}{\partial x_{1}}  & \frac{\partial F_{1}^{x}}{\partial x_{2}} 
			& \frac{\partial F_{1}^{x}}{\partial y_{1}}
			& \frac{\partial F_{1}^{x}}{\partial y_{2}} & \frac{\partial F_{1}^{x}}{\partial \theta_{1}}
			& \frac{\partial F_{1}^{x}}{\partial \theta_{2}} \\\\
			\frac{\partial F_{2}^{x}}{\partial x_{1}}  & \frac{\partial F_{2}^{x}}{\partial x_{2}} 
			& \frac{\partial F_{2}^{x}}{\partial y_{1}}
			& \frac{\partial F_{2}^{x}}{\partial y_{2}} & \frac{\partial F_{2}^{x}}{\partial \theta_{1}}
			& \frac{\partial F_{2}^{x}}{\partial \theta_{2}} \\\\
			\frac{\partial F_{1}^{y}}{\partial x_{1}}  & \frac{\partial F_{1}^{y}}{\partial x_{2}} 
			& \frac{\partial F_{1}^{y}}{\partial y_{1}}
			& \frac{\partial F_{1}^{y}}{\partial y_{2}} & \frac{\partial F_{1}^{y}}{\partial \theta_{1}}
			& \frac{\partial F_{1}^{y}}{\partial \theta_{2}} \\\\
			\frac{\partial F_{2}^{y}}{\partial x_{1}}  & \frac{\partial F_{2}^{y}}{\partial x_{2}} 
			& \frac{\partial F_{2}^{y}}{\partial y_{1}}
			& \frac{\partial F_{2}^{y}}{\partial y_{2}} & \frac{\partial F_{2}^{y}}{\partial \theta_{1}}
			& \frac{\partial F_{2}^{y}}{\partial \theta_{2}} \\\\
			\frac{\partial T^{z}_{1}}{\partial x_{1}}  & \frac{\partial T^{z}_{1}}{\partial x_{2}} 
			& \frac{\partial T^{z}_{1}}{\partial y_{1}}
			& \frac{\partial T^{z}_{1}}{\partial y_{2}} & \frac{\partial T^{z}_{1}}{\partial \theta_{1}}
			& \frac{\partial T^{z}_{1}}{\partial \theta_{2}} \\\\
			\frac{\partial T^{z}_{2}}{\partial x_{1}}  & \frac{\partial T^{z}_{2}}{\partial x_{2}} 
			& \frac{\partial T^{z}_{2}}{\partial y_{1}}
			& \frac{\partial T^{z}_{2}}{\partial y_{2}} & \frac{\partial T^{z}_{2}}{\partial \theta_{1}}
			& \frac{\partial T^{z}_{2}}{\partial \theta_{2}} \\\\
		\end{bmatrix}
	\end{math}
\end{center}

\bibliography{ALL.remote}

\end{document}